# Wi-attack: Cross-technology Impersonation Attack against iBeacon Services

Xin Na, Xiuzhen Guo, Yuan He, Rui Xi
School of Software & BNRist, Tsinghua University, China
nx20@mails.tsinghua.edu.cn, guoxiuzhen94@gmail.com, he@greenorbs.com, ruix.ryan@gmail.com

***Abstract*—iBeacon protocol is widely deployed to provide location-based services. By receiving its BLE advertisements, nearby devices can estimate the proximity to the iBeacon or calculate indoor positions. However, the open nature of these advertisements brings vulnerability to impersonation attacks. Such attacks could lead to spam, unreliable positioning, and even security breaches. In this paper, we propose Wi-attack, revealing the feasibility of using WiFi devices to conduct impersonation attacks on iBeacon services. Different from impersonation attacks using BLE compatible hardware, Wi-attack is not restricted by broadcasting intervals and is able to impersonate multiple iBeacons at the same time. Effective attacks can be launched on iBeacon services without modifications to WiFi hardware or firmware. To enable direct communication from WiFi to BLE, we use the digital emulation technique of cross technology communication. To enhance the packet reception along with its stability, we add redundant packets to eliminate cyclic prefix error entirely. The emulation provides an iBeacon packet reception rate up to 66.2%. We conduct attacks on three iBeacon services scenarios, point deployment, multilateration, and fingerprint-based localization. The evaluation results show that Wi-attack can bring an average distance error of more than 20 meters on fingerprint-based localization using only 3 APs.**

## I. Introduction

Over the last decade, we have witnessed the explosive growth of the Bluetooth Low Energy (BLE) technology [1]–[4], which can be used to achieve communication and locate everyday objects around us. Moreover, top technology companies like Apple, have invested much research into this field through iBeacons [5]. Since Apple introduced the protocol in 2013, various vendors have manufactured iBeacon compatible hardware, called iBeacons. With the help of these iBeacons, services such as "check-in service" or "notifications on newly released products" [6] are provided in the mall or sports hall.

Nowadays, iBeacon is widely deployed to provide location services in indoor positioning systems (IPS) based on BLE [3]. Specifically, an iBeacon periodically broadcasts BLE frames containing a received signal strength indicator (RSSI) value at one meter from the iBeacon station. BLE-equipped smartphones can detect them and estimate the distance to the iBeacon station. For example, to create a more engaging shopping experience, item information will appear on a user's phone when the user nears a store. However, the estimation accuracy can not be guaranteed due to channel noise, signal attenuation, and multipath propagation. Therefore, different from the accurate tracking or localization method [7], [8], iPhone provides only proximity to the iBeacon, (e.g., immediate, near, far). Another method of building a fingerprint database for each point of interest (PoI) provides relatively accurate and reliable positioning. It leverages the RSS information from multiple iBeacons and achieves a meter-level accuracy [3].

Whereas the open nature of iBeacon advertisement brings vulnerability against impersonation attacks [9]. Since iBeacon broadcasts its unique ID and transmission power publicly, any third party can obtain the content using a cellphone. This gives a chance for anyone to make a replica of the target iBeacon. The attacker utilizes a BLE compatible hardware to mimic an iBeacon device by broadcasting the same packet. It leads to severe consequences such as spam [6], misdirection [9], or even security breaches [10]. For example, for indoor localization systems based on iBeacon, fake iBeacons can bring higher distance error or even mislead users.

In this paper, we propose Wi-attack, which leverages the wide-deployed WiFi devices (such as WiFi APs) to conduct poisonous impersonation attacks towards iBeacon services. The attacker only needs to obtain the ID of the iBeacon with a cellphone first. Then, he can launch the attack by using WiFi APs under his control. The WiFi packets transmitted by the WiFi AP are regarded as legitimate BLE packets on the iBeacon devices and normal WiFi packets on the WiFi devices, respectively. In this way, Wi-attack mimics iBeacons' behavior and is considered to be communicating with another WiFi device normally at the same time. This makes it hard to be detected. Moreover, Wi-attack has the ability to be implemented on any WiFi device with no modification on hardware or firmware, which provides high attack efficiency with extremely low overhead.

In order to perform impersonation attacks on iBeacon services with WiFi devices, we face several challenges. Firstly, due to the strict decoding mechanism of the BLE receiver, any bit error will immediately lead to a reception failure. The existing signal emulation method [11]–[13], however, relies heavily on the error tolerance technique (DSSS) in ZigBee. They cannot provide a Wi-Fi to BLE link due to the inherent high emulation errors. Therefore, a high-precision emulation method is necessary. Secondly, considering the practical application, our method needs to be able to attack effectively under different iBeacon scenarios.

To solve the above challenges, we develop a novel cross-technology communication (CTC) system called Wibeacon. Then, we provide methods of attacking existing iBeacon services by using wide deployed WiFi APs to transmit fake iBeacon packets. The contributions of this work are as follows:

- By analyzing the transmission and reception of the iBeacon packet as well as the working mechanism of iBeacons, we reveal the feasibility of using the WiFi device to conduct impersonation attacks on iBeacon services. The WiFi AP is able to transmit fake but legitimate iBeacon packets without any hardware modification.
- To enhance the attack efficiency, we propose Wibeacon to improve the packet reception rate (PRR). Compared to the two existing approaches WEBee [12] and WIDE [13], Wibeacon improves the PRR of the emulated iBeacon packet to 66.2%, while the value is 0% on both WEBee and WIDE. We also present the attack methods under three different scenarios.
- We implement Wi-attack and evaluate the performance of impersonation attacks on the real-world deployment of iBeacon services. The results show that Wi-attack can bring an average distance error of up to 20 meters in a common fingerprint-based localization system.

The rest of this paper is organized as follows. We present the motivation of this work in Section II. Section III introduces our threat model. Section IV presents the main design of Wi-attack. We show the evaluation result of our system in Section V. Section VI discusses related works. In Section VII, we make some discussion. We conclude this work in Section VIII.

## II. Motivation

In this section, we introduce existing attack approaches against iBeacon and their limitations. Further, we analyze what it takes to perform an effective attack.

### A. Existing Approaches

There are two common types of attacks on iBeacon services, namely the jamming attack and impersonation attack.

For the jamming attack, the attacker utilizes a jammer device (usually a WiFi device) to prevent BLE communication by occupying the channel or colliding their packets [14], [15]. It's easy to use a commodity WiFi device to launch such attacks because it only requires modifications to the chip's parameters. To defend against such an attack, many works have been proposed to recover the collided packets or protect legitimate communications [16]–[18]. Moreover, such methods have limited ability as they can not launch advanced attacks such as providing a false location.

For the impersonation attack, the attacker first sniffs the UUID (Universally Unique Identifier) of an iBeacon. Then he uses a BLE device to mimic an iBeacon by broadcasting the same advertisement. This fake iBeacon could send spam to the user or introduce a higher distance error to the localization. However, the ability of these fake iBeacons is limited. On the one hand, many fake iBeacons need to be deployed first to mislead the user effectively. On the other hand, the broadcasting interval of the BLE peripheral devices restrains the attacker from sending out advertisements as desired. The small region of the fake iBeacons also limits the influence.

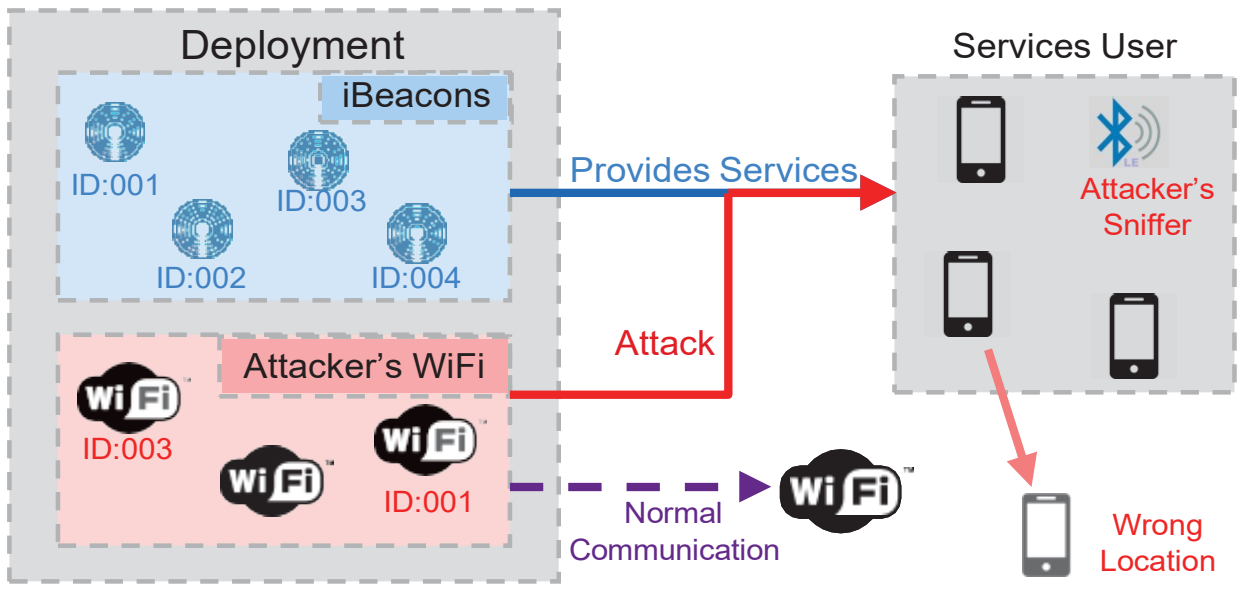

Fig. 1: The illustration of attack model: Information of the deployed iBeacons are obtained by a sniffer. Then, impersonation attack is launched to alter the localization result of the user. The attacker's WiFi devices is regarded to be communicating with legitimate WiFi during the attack.

### B. Analysis

Since the above two attack methods are both limited, we now explore how to conduct a truly effective and poisonous attack while being hard to be detected.

There are three values used to perform operations in the iBeacon service: the unique ID of the iBeacon, the reference TX transmission power, and the RSSI value. A BLE device uses unique IDs to identify different iBeacons and measures the strength of the received signal to obtain RSSI. Comparing this value with the reference TX power value, the device is able to estimate its distance to the iBeacon.

Therefore, manipulating these values is the key to effective attacks. The only way to achieve this is by impersonating trusted iBeacons because the receiver ignores packets containing unknown IDs. In order to provide such attacks with high effectiveness, a few features are required: 1) low implementation overhead with the ability to impersonate multiple iBeacons at the same time, 2) the ability to alter the TX power reference freely, 3) the ability to manipulate the RSSI value experienced by the user in a wide range, especially for fingerprint-based positioning system, and 4) high concealment.

The method we propose, **Wi-attack**, has all the above features. 1) Any iBeacon can be impersonated by only modifying WiFi's payload. Moreover, with no intervals required between broadcasting, it can impersonate multiple iBeacons at the same time. 2) The strong signal strength provides a wider attack range. It also enables us to manipulate the RSSI in a wide range. By carefully choosing the broadcasting scheme on the WiFi AP, effective and advanced attacks can be achieved on iBeacon services. 3) These attacks are hard to be detected because the WiFi device is transmitting legitimate packets and is considered to be communicating with other devices normally.

## III. Threat Model

We now introduce the threat model of Wi-attack. Wi-attack aims to conduct attacks on common deployments of iBeacon, as depicted in Fig. 1. One or more iBeacons are deployed in the area to provide services such as pushing notifications or localization. The BLE device receives advertisements and performs corresponding actions. The attacker first uses a BLE compatible device, such as a cellphone, to collect IDs of iBeacons in the area. After obtaining these IDs, the attacker decides the target iBeacons to be attacked. By modifying the payload, the WiFi device can transmit BLE advertisement packets to the victims. Then, the attacker sets the WiFi device's transmission power according to his aim to initiate the attack. Because the WiFi device sends out legitimate WiFi packets and impersonates normal iBeacons, it is difficult for others to detect and defend against such an attack.

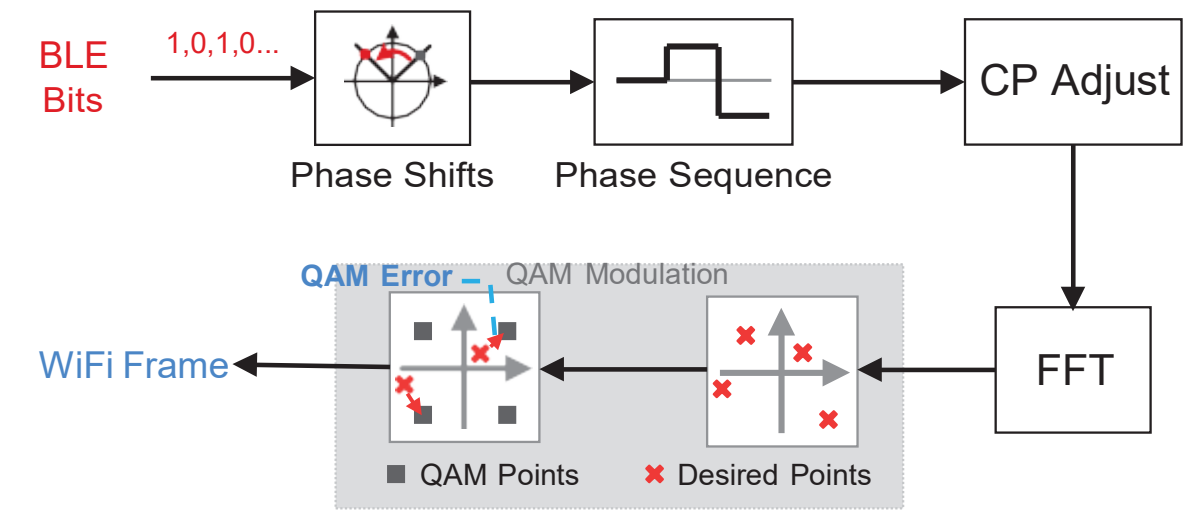

Fig. 2: The workflow of digital emulation

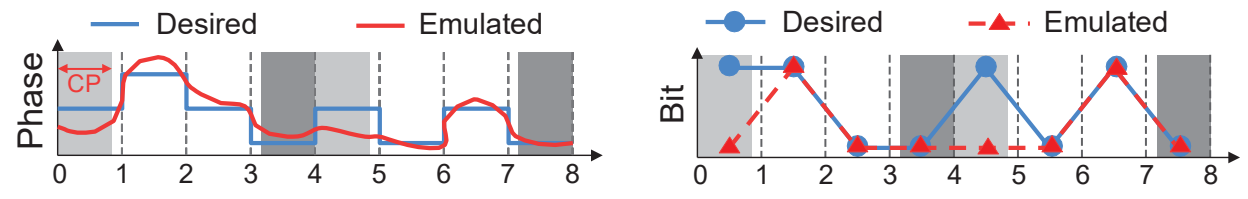

(b) Emulation result

Fig. 3: An example of digital emulation

## IV. Design

This section introduces the design of our system. Based on the digital emulation method proposed in [13], we present how to send out BLE packets with WiFi first. Then, we introduce attacks under three typical scenarios. Finally, to make the attack more effective, we introduce an enhanced emulation method from WiFi to iBeacon.

### A. Wibeacon Generation

In order to enable direct communication from WiFi to BLE so that the advertisement packets could be received, we design a CTC system called **Wibeacon**. Based on the digital emulation technique of CTC [13], Wibeacon further splits each phase ladder into fine-grained ones. It emulates legitimate iBeacon packets simply by modifying the payload of the WiFi frame.

BLE uses Gauss Frequency Shift Keying (GFSK) to modulate data bits with frequency shifts of the signal. It can be converted to a phase shift modulation:

$$s(t) = A\cos(2\pi(f + \Delta f)t) = A\cos(2\pi f t + \phi(t)) \quad (1)$$

where the phase shift "+" and "-" of two continuous sampling points in $\phi(t)$ correspond to bit "1" and "0". These phase shifts are constructed to a ladder-shaped phase sequence and input to the digital emulation process shown in Fig. 2. The WiFi frame is then obtained by reversing its modulation procedure including copying the cyclic prefix and finding the closest constellation points in Quadrature Amplitude Modulation (QAM). One example result is shown in Fig. 3. For the 8 bits signal that lasts for 8 $\mu$s, we use 2 WiFi symbols. The last 0.8$\mu$s of each symbol is exactly the same as the first 0.8$\mu$s, which is the cyclic prefix (CP) in WiFi. The existence of CP brings errors during signal emulation. Fig. 2 shows another source of error known as QAM error. In Fig. 3, two bits of error occur in the digital emulation, which is unacceptable because any bit error leads to packet reception failure in BLE.

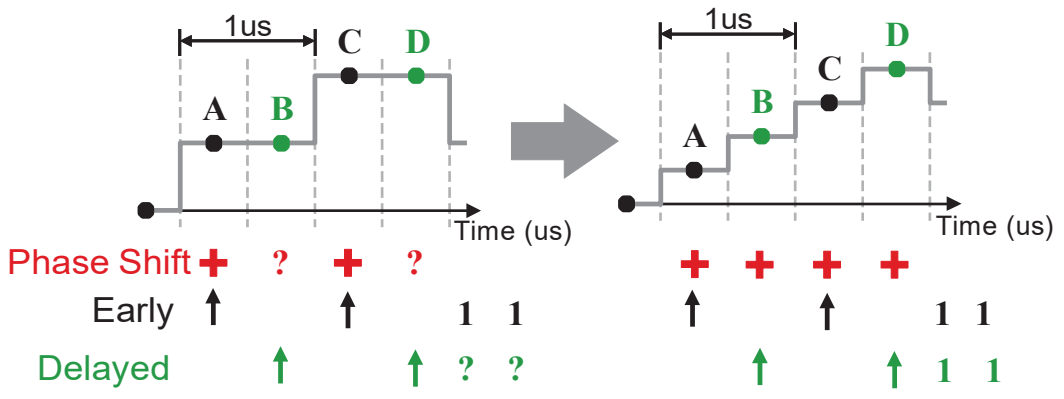

Fig. 4: Split the ladder shift sequence

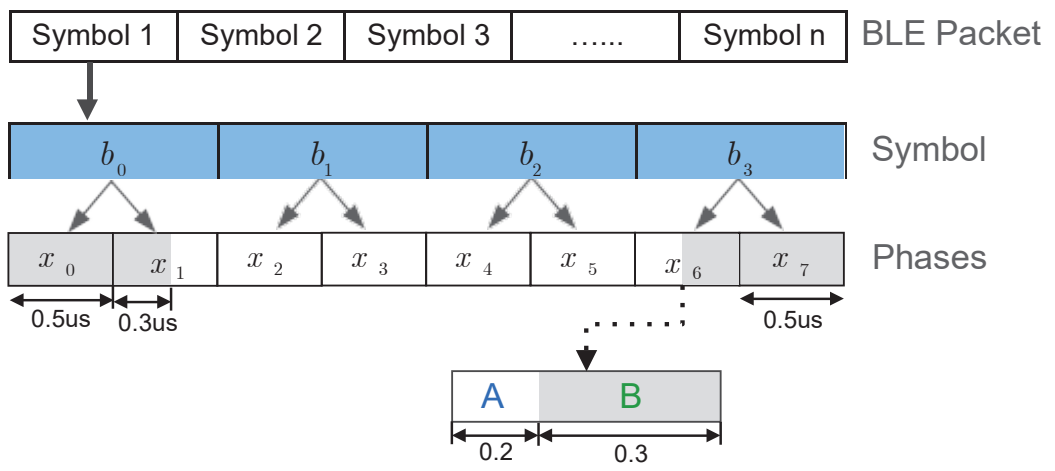

Fig. 5: Phase sequence of BLE Packet

To address the above question, we find that BLE uses downsampling for decoding where every two sampling points are used to decode one bit in every 1$\mu$s. The downsampling leads to two different decoding cases called "early decoding" and "delayed decoding". Therefore, we split each 1-$\mu$s-long ladder into two fine-grained ladders as shown in Fig. 4. This makes the result of "delayed decoding" also correct and is the same with "early decoding". Taken the more refined sequence as the input, WiFi is able to emulate in a more precise manner. As shown in Fig. 5, every 4-$\mu$s-long WiFi symbol corresponds to a 4-bit BLE symbol, which is formed by 8 phase ladders. However, to meet the requirement of CP in WiFi, part of phase $x_6$ and the entire $x_7$ cannot be altered freely. They must be exactly the same as the first 0.8$\mu$s (phase $x_0$ and part of $x_1$). To analyze the influence on the decoding result, we build a probability model.

BLE samples in every 0.5$\mu$s, which means there exists one sampling point in every phase ladder in Fig. 5. Also, the position of each point on every ladder is the same. We denote the possibility that the sampling point falls into A section of phase $x_6$ (the first 0.2$\mu$s) as $P(A)$ and B section as $P(B)$. Then, $P(A) = 40\%$ and $P(B) = 60\%$. We use $P(W|A)$ and $P(W|B)$ to represent the error probability under situations A and B, respectively. The probability of a correctly decoded BLE symbol is:

$$P = 1 - P(A)P(W|A) - P(B)P(W|B) \quad (2)$$

For different types of BLE symbols, the decoding rate given by this model is different. For the symbols that satisfy $b_0 = b_3$, including "0000", "0010", "1011", and etc. As an example shown in Fig. 6, its phase shift of $b_3$ is the same as $b_0$. No matter the sampling point is in case A (the blue triangle) or case B (the green square), the decoding result is always correct.

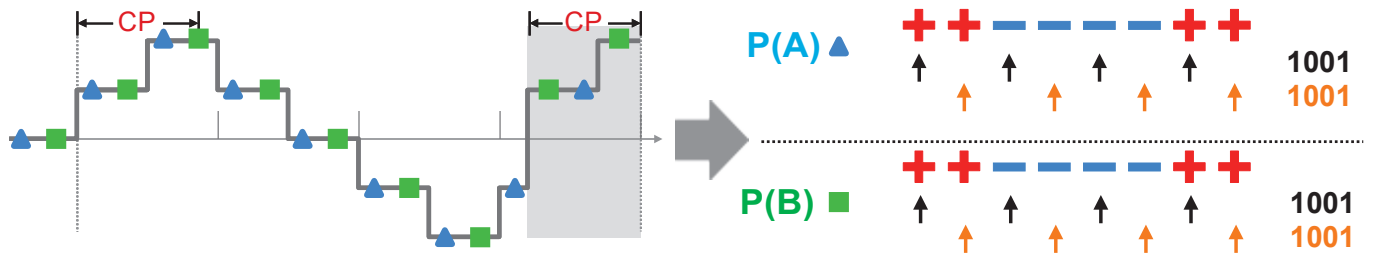

Fig. 6: Decoding phase sequence of symbol "1001"

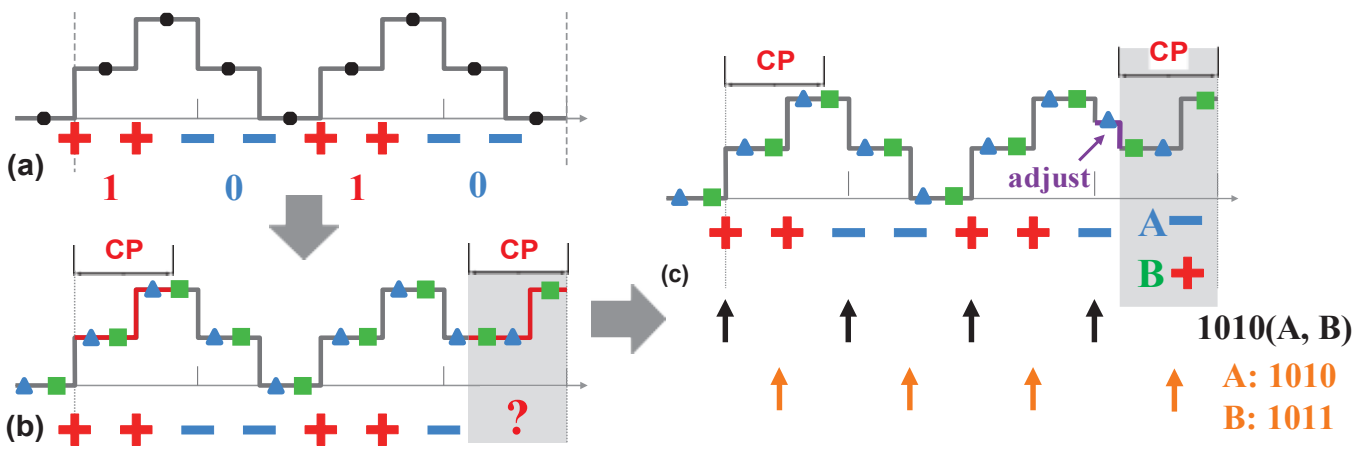

Fig. 7: Decoding phase sequence of symbol "1010"

However, for the $b_0 = b_3$ symbols, including "0001", "0011", "1010", and etc. As the example of symbol "1010" is shown in Fig. 7, $b_3$ will not be obtained correctly under delayed decoding for case A and case B. Neither the phase shift meets the requirement of bit 0, a negative shift. Therefore, we have $P(W|A) = P(W|B) = 50\%$ because the probability of early decoding case is 50%. To improve this, we make an adjustment to the phase ladder in 3.0-3.2 $\mu$s as shown in Fig. 7(c). When the sampling point is in case A, the result will be correct under delayed decoding after the adjustment. In this way, we improve $P(W|A)$ to 0. Now the probability of correct decoding is $P = 1 - 0.6 \times 0.5 = 70\%$.

As it's not possible to avoid using $b_0 = b_3$ symbols to form an iBeacon packet. We now provide the ability to transmit iBeacon packets with WiFi under a PRR of 70% without considering QAM error. Later, we show the actual PRR is 34.4% in this case.

### B. Attack Initiation

After having the ability to emulate iBeacon packets with WiFi devices, we introduce preparations before the attack. This includes information gathering, channel coordination, and decisions on setting the parameters.

For information gathering, the attacker needs to grasp the iBeacon deployment plan for a target region. This can be done by using a sniffer such as a cellphone. The attacker first obtains the iBeacon IDs and then matches the received IDs with the specific deployed iBeacons through observing the RSS value while approaching it. If he desires to launch advanced attacks on fingerprint-based positioning systems, such as misleading the user from one spot to the other. More information needs to be gathered such as the whole RSS fingerprint database. The authors in [9] have shown this is achievable.

Then, the attacker needs to coordinate the channels to transmit emulated packets. There are three broadcasting channels for BLE, namely 37, 38, and 39, located on 2402MHz, 2426MHz, and 2480MHz, respectively. Each of them has a 2MHz bandwidth. When a BLE device scans for advertisements, it sweeps from all three channels and scans each of them for a short period as shown in Fig. 8. To address this, iBeacon transmits one packet on all three channels every time. Therefore, each emulated packet also needs to be transmitted on multiple channels. As shown in Fig. 9, 802.11 g/n/ac and BLE share the spectrum between 2.402 to 2.480 GHz. WiFi channel 4 and 13 are able to cover the entire spectrum of BLE channel 38 and 39. But channel 1 overlaps only 1MHz with BLE channel 37, which is not sufficient for packet emulation. In this way, Wi-attack can only transmit advertisements on BLE channel 38 and 39. This coverage is enough in most cases because the BLE device only scans on each channel for a short period (eg. tens of milliseconds).

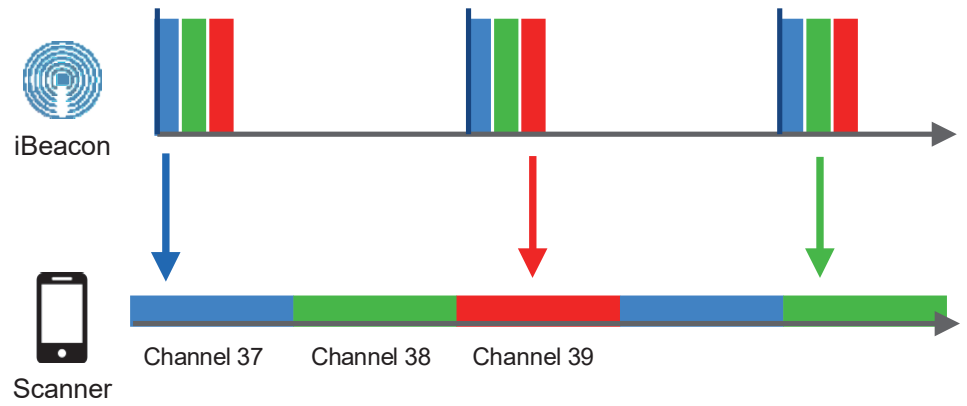

Fig. 8: The broadcasting and scanning procedure of BLE

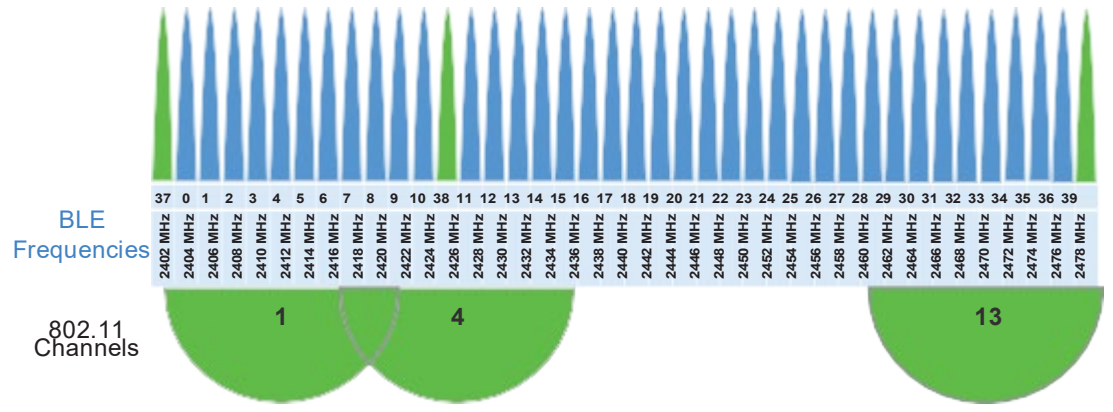

Fig. 9: Channel coordination between WiFi and iBeacon

Finally, the attacker needs to decide three parameters to launch certain attacks: 1) the ID of the iBeacon to impersonate, 2) the TX power value, and 3) the actual transmitting power of the WiFi AP. Note that the attacker can impersonate multiple iBeacons by broadcasting their packets at the same time. The TX power value is contained within the packet as a fake power reference while the actual transmitting power of WiFi AP can be set arbitrarily. By manipulating these values, the attacker is able to tell the device a false distance or location.

### C. Attack Scenarios

We consider three typical scenarios of iBeacon deployments for the attack to take place.

**Point deployment** is the simplest deployment plan. One iBeacon is deployed to provide services such as "push notifications nearby". The device uses the measured RSS value and the TX power value to approximate the distance to the iBeacon. As shown in Fig. 10, Wi-attack interferes with the distance estimation process between iBeacon and its customer. The distance between the device with the iBeacon is given by:

$$d = 10^{\frac{P-S}{10n}} \tag{3}$$

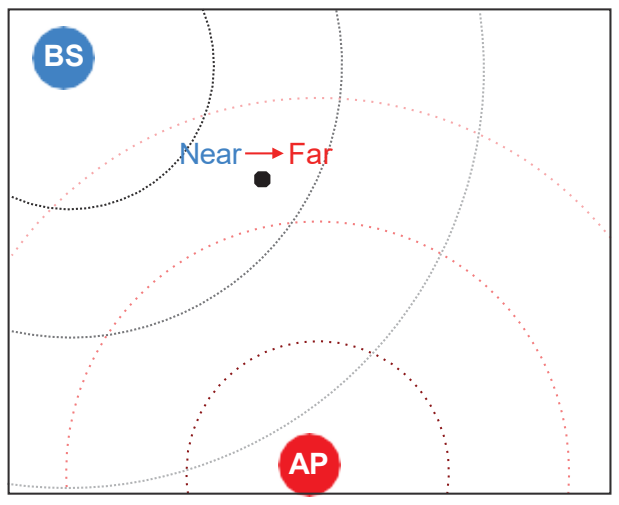

Fig. 10: Attack on point deployment scenario

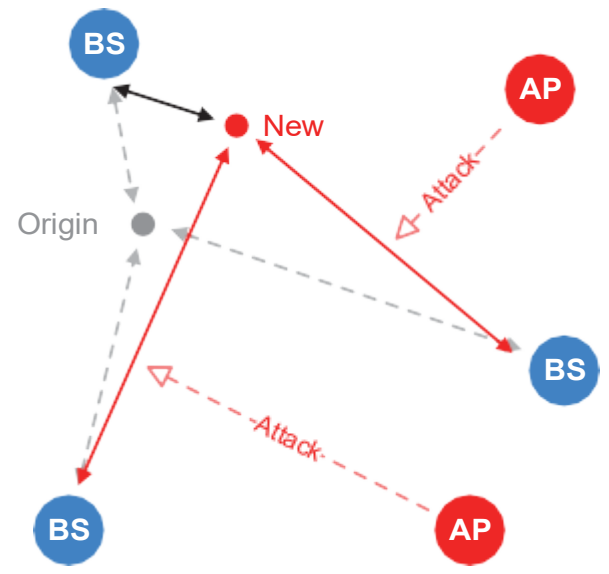

Fig. 11: Attack on multilateral localization

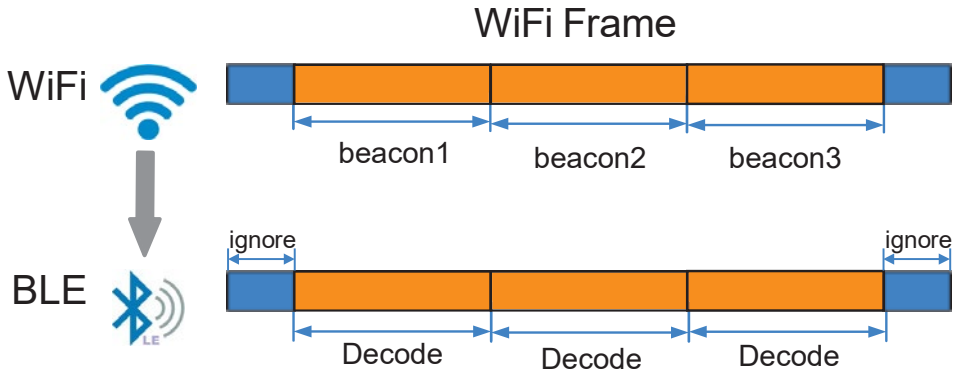

Fig. 12: Supplementary packets design in enhanced Wibeacon

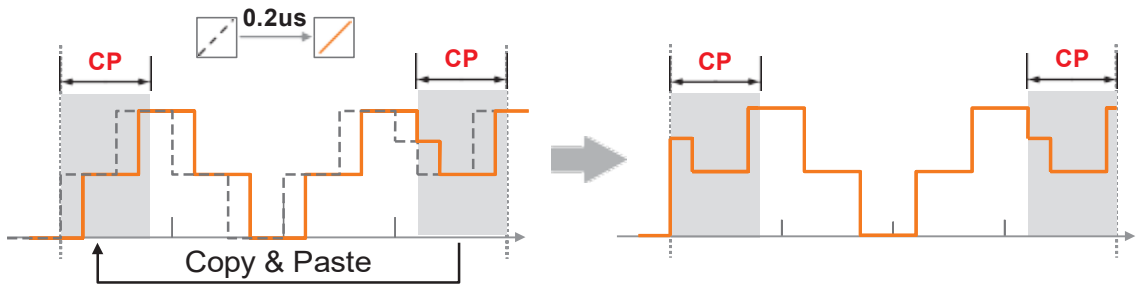

Fig. 13: Add a delay to the original phase sequence

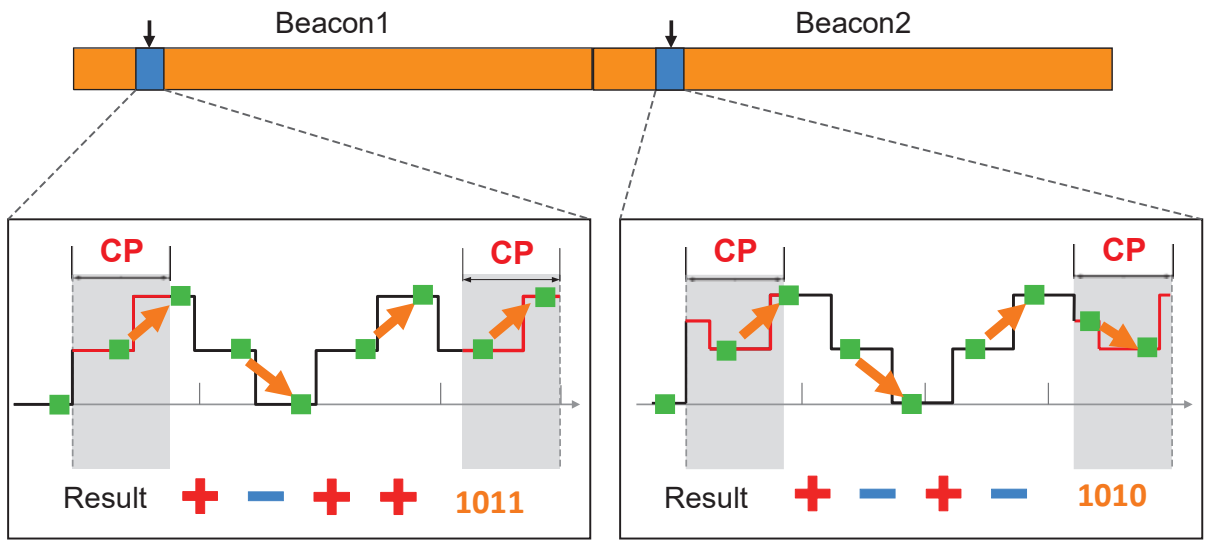

Fig. 14: Decoding result with the first supplementary packet

where $p$ stands for the TX power reference, $s$ stands for the RSS value measured by the device. $n$ is a signal propagation factor which is a constant.

We show that in order to launch an effective attack, the attacker needs to set a high transmit power along with a high reference power. Firstly, if the attacker desires to affect as many devices as possible, he needs to increase the range of WiFi AP by increasing the actual transmit power. Secondly, the attacker wants the customer to believe he is far from the shop when he is actually near. This is achieved by setting a high fake TX power reference. Because according to Eq.3, a longer distance is obtained when the $p$ $s$ value is large.

**Multilateral localization** uses multiple measured distances to calculate a relative position as shown in Fig. 11. The attacker attempts to tamper with the localization result by affecting the distance measurement as much as possible and as far as possible. Here, we show that the attacker is able to provide any distance to any iBeacon.

We define the true TX power reference which is the RSS value one meter away as $p_0$, and define the fake power reference we inject to the packet as $p_f$ . When the true distance between the user and the attacker's AP is $d_0$, the RSS value measured by the device can be represented as $s = p_0 \; 10n \lg(d_0)$. The distance we make the device to believe would be:

$$d_f = 10^{\frac{p_f - s}{10n}} = \frac{d_0}{d_{f0}} \tag{4}$$

where $d_{f0}$ stands for the distance where the measured RSS value equals the injected $p_f$ . As we can inject the RSS value of any distance in range, the fake distance can be of any value to result in a faulty position.

**Fingerprint-based localization** achieves a more precise localization by storing all RSS values for the points of interest as a database. The location is calculated using the database and the collected RSS value from all iBeacons.

Many works achieve effective RSS attacks on fingerprint-based systems [9], [19]. All of them can be used in our case. We also highlight that using WiFi AP to conduct these attacks on iBeacons has several advantages. 1) Having higher maximum transmit power, WiFi AP can provide a wider RSS range that further increase the attacker's ability. 2) A WiFi AP can impersonate multiple iBeacons at the same time without its interval restriction, which reduces the AP number needed for an advanced attack. 3) With much more multipath propagation, the measured RSS value is more unstable over time, which brings higher dynamic distance error for the returned location at one spot.

In the practical attack scenario, one important problem is to decide the specific iBeacon device for each AP to impersonate. A reasonable assignment can lead to a higher distance error under attack, such as choosing an iBeacon that is in a faraway location. Since the WiFi APs are widely deployed, the attacker is always able to obtain an effective impersonation plan. It is worth noticing that we leave the problem of obtaining such a plan in our future work. We conduct a series of experiments to evaluate the performance of using WiFi AP to attack fingerprint-based systems, which can be seen in §V-E.

## D. *Attack Enhancement*

Although the above method is able to launch impersonation attacks, we find that the effectiveness is low due to the low PRR Wibeacon provides. BLE devices are always able to receive the "good" packets from iBeacons but only successfully decodes our emulated packets with low accuracy. Therefore, we find it critical to enhance the packet reception.

The 70% decoding rate provided by Wibeacon will be reduced to around 30% after QAM emulation in Fig. 2. However, the error brought by QAM is inherent and fixed in emulation. We still seek to eliminate the error brought by CP. We notice that the missing 30% theoretical PRR is caused by failure to decode $b_0 = b_3$ symbols under delayed decoding case when the sampling point is located in segment

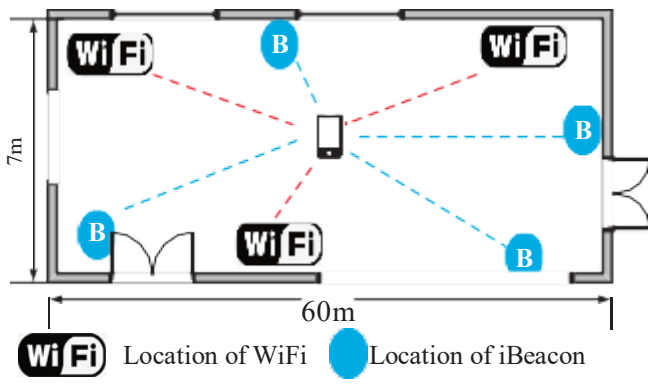

Fig. 15: The floor Plan for Multilateral Localization

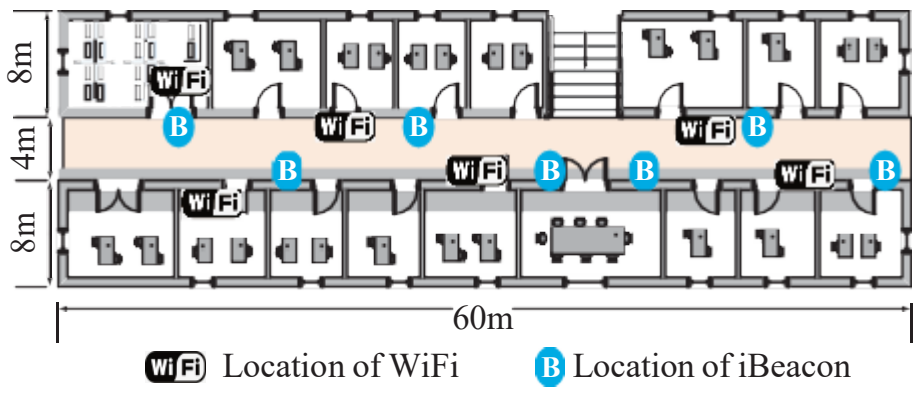

Fig. 16: The floor Plan for Fingerprint Localization

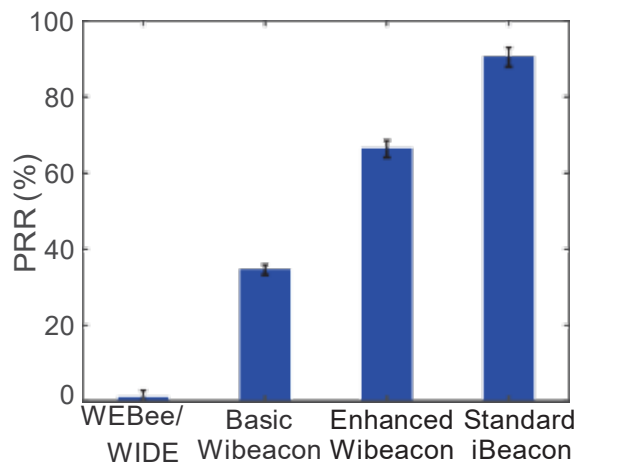

Fig. 17: Emulation method performance comparison

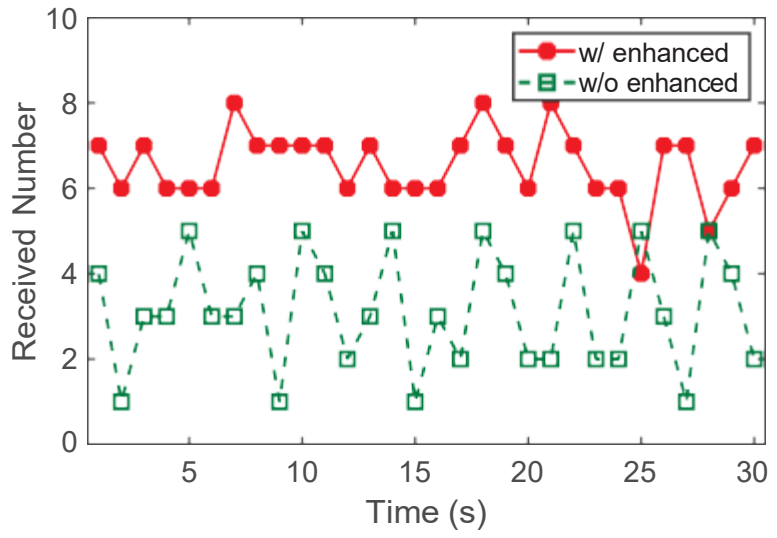

Fig. 18: Comparison of packet reception stability

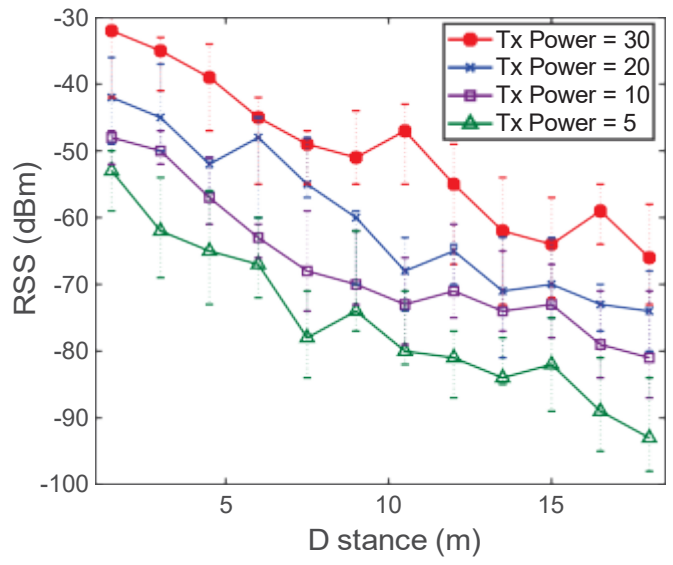

Fig. 19: RSSI range provided by impersonate iBeacon

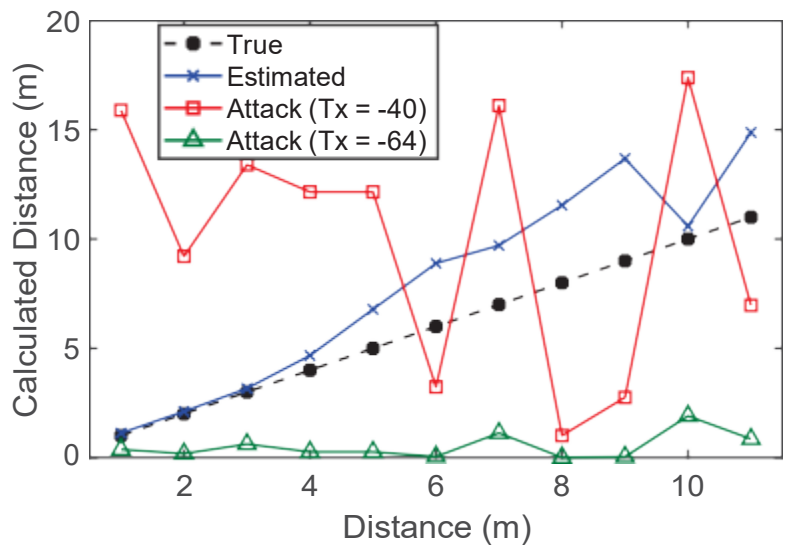

Fig. 20: Attack on point deployment: estimated distance

B (0.2-0.5$\mu$s in phase $x_6$) in Fig. 5. But the decoding result is always correct when the sampling point is located in segment A. We attempt to drag the sampling points out of segment B to segment A by providing delays in the decoding process.

Considering that the original phase sequence can no longer be changed, we add supplementary packets after the original as shown in Fig. 12. The latter packets contain the same information but have a built-in delay in the time domain. When the decoding fails for *beacon*1, *beacon*2 delays 0.2 $\mu$s to let the sampling point jump out of segment B and jump into segment A. The reason for using two supplementary packets is that segment A is shorter than segment B and cannot cover all sampling point positions after the delay.

We take the symbol "1010" in *beacon*2 as an example to show the new phase generation method. To provide such a built-in delay, we move the phase sequence to the right by 0.2$\mu$s as shown in Fig. 13. To meet the requirement of CP, we replace the sequence in $[0, 0.2]\mu$s by copying the sequence in $[3.2, 3.4]\mu$s. The phase sequence in *beacon*3 can be generated in the same way by moving 0.3$\mu$s.

As shown in Fig. 14, the error once brought by fragment B and delayed decoding is now eliminated in *beacon*2. But it's not enough using only *beacon*2. If the sampling point is in $[0.4, 0.5]\mu$s in *beacon*1, it will be dragged to $[0.3, 0.4]\mu$s by *beacon*2, which is still in segment B. Therefore, *beacon*3 is needed to resolve this. After this, we obtain a 100% decoding accuracy by eliminating all CP errors. Note that the decoding accuracy here is only for square wave signal including CP. The emulation error brought by QAM modulation is not considered, and it will result in a PRR lower than excepted. Also, bringing the supplementary packets will not reduce the attacking efficiency. Because the packet period is less than 1 *ms* which is neglectable compared to packet interval.

## V. Evaluation

### A. Implementation

We implement Wi-attack on the USRP platform and use cellphones to evaluate its performance. We use a USRP N210 device with 802.11 g/n PHY to transmit Wibeacon packets. For the standard iBeacon services, we use HackRF devices with BLE PHY. We set the advertisement interval for both devices to be 100 ms. The reason to use USRP and HackRF is that we are able to obtain the packet level information and change the broadcast schemes easily with them. Wi-attack can be realized by many commodity WiFi devices (e.g. Atheros AR9485, AR5112, and AR2425) since it only modifies WiFi's payload. We set the channels of the WiFi transmitter as shown in Fig. 9. To receive iBeacon packets and measure its RSS, we use a commodity BLE chip CC2650, and an iPhone Xs Max running the iOS version 12.5.

We deploy iBeacons and WiFi APs in two environments as shown in Fig. 15 and 16 to implement multilateral localization and fingerprint-based localization. We implement multilateral localization so that it finds the optimal coordinates by minimizing the error function. For fingerprint-based localization, we implement a weighted-kNN matching algorithm presented in [3].

### B. Wibeacon Benchmark

*1) Performance comparison:* First, we evaluate the effectiveness of the Wibeacon packets compared to previous CTC methods. We compare the performance with WEBee [12] and WIDE [13] by sending 1000 packets under the same channel and calculate the packet reception ratio (PRR), as shown in Fig. 17. The basic Wibeacon (with fine-grained phase ladder) generated with digital emulation reaches a PRR of 34.4% while WEBee and WIDE packets can hardly be received. However, as we have mentioned that Wibeacon can only be transmitted on BLE channel 38 and 39, the PRR for iPhone and other BLE devices reduces to about two-thirds of the original. Hence, we find it unacceptable and propose the enhanced Wibeacon that has a PRR of 66.2%. Enhanced Wibeacon resolves errors brought by WiFi CP entirely, but the inherent QAM error prevents it from reaching the level of standard iBeacon.

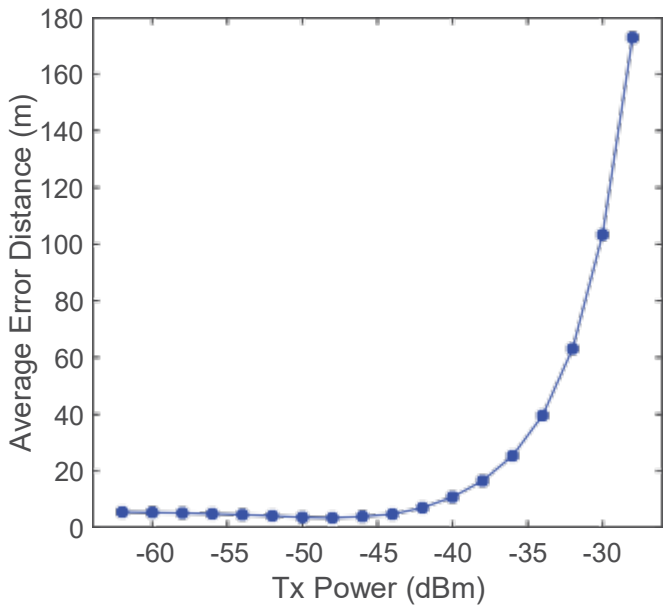

Fig. 21: Impact of Tx Power setting on distance error

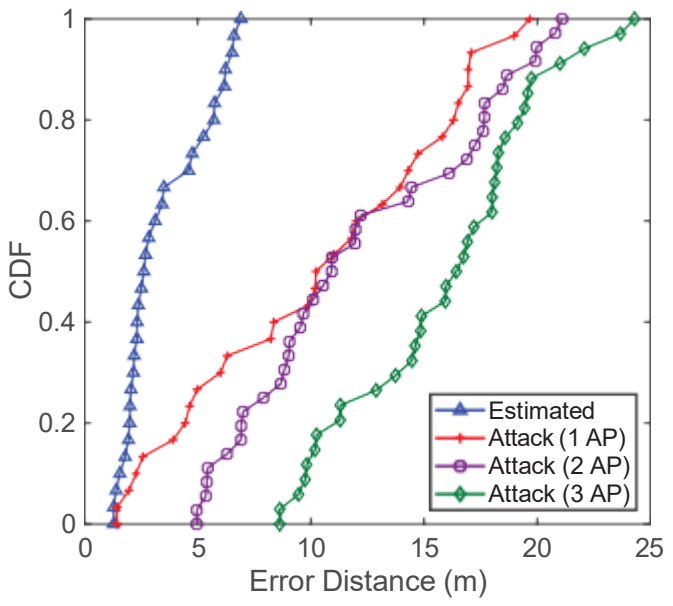

Fig. 22: Attack on Multilateral Localization: error distance

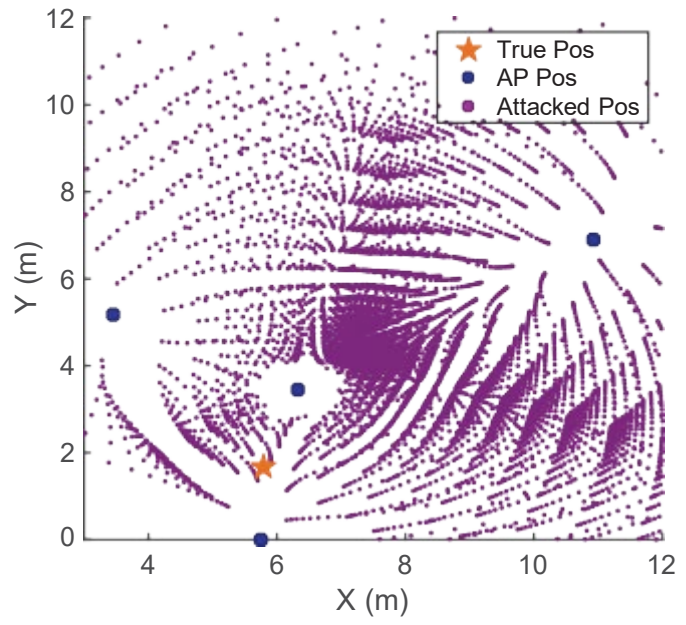

Fig. 23: Case study: impact of Tx power of localization result

*2) Receiving stability:* Another problem with Wibeacon is that it provides a very unstable reception. To compare the stability between basic and enhanced Wibeacon, we transmit the emulated packets for 30 seconds using 100 ms intervals. The number of received packets in every second is shown in Fig. 18. While the number jumps between 1 and 5 with Wibeacon, the reception of enhanced Wibeacon is much more stable over time.

### C. Point Deployment

*1) RSSI Range:* We first verify whether fake iBeacon using WiFi provides a wider range of RSSI experienced by the user. We place the fake iBeacon in an indoor area and set the transmission power of the fake iBeacon to different levels. Fig. 19 shows the average RSS under these levels. We see that a fake iBeacon using WiFi provides an RSSI ranging from -90 to -30, while the range on a normal iBeacon is from -95 to -65. In this way, the attacker is able to manipulate the RSSI experienced by the user in two times wider range.

*2) Attack performance:* We now evaluate the attack performance under point deployment. We place a standard iBeacon and a fake iBeacon in a wide indoor area. We set the WiFi's transmission power to a high level as explained in §IV-C. The TX power reference in the standard iBeacon is -64, which is the RSSI measured one meter away. For the fake iBeacon, we use two values: -40 and -64. -40 stands for a normal reference power for WiFi AP, while -64 is a lower power reference. We randomly choose positions and estimate the distance to the iBeacon. Fig. 20 shows the estimated distance under different settings. With a high reference power such as -40, the estimated results are greatly affected under attack. However, the reference power -64 is close to or even lower than the measured RSSI. Therefore, the estimated result is always low.

*3) Impact of TX power setting:* In the previous experiment, the TX power reference is a fixed value, now we evaluate how it affects the estimation error. By setting different TX reference power, we calculate the average distance error under attack. As shown in Fig. 21, the estimation error increases exponentially with TX power reference. The reason is that the distance calculated in Eq.3 increases exponentially with value $p$ s. This suggests that using a higher reference power is a better option. However, setting the value too high could expose the attacker. Therefore, we recommend setting the value to around -40, which always brings a few meters error.

### D. Multilateral Localization

In this section, we conduct attacks on multilateral localization systems. Note that due to the existence of multipath propagation, the multilateral localization system based on iBeacons usually has suboptimal performance. Therefore, we choose a wide indoor area with few barriers. We place the iBeacons and fake iBeacons as shown in Fig. 15.

*1) Attack performance:* In this experiment, we evaluate the attack performance while the TX Power reference is still fixed to -40. Fig. 22 shows the CDF of error distance with different numbers of APs used. With more fake iBeacons engaged to conduct the attack, the distance error increases significantly. The result also shows that using only one fake iBeacon is enough to conduct an effective attack. This is because multilateration relies heavily on each estimated distance. If one of them is compromised, the estimated location will be greatly affected.

*2) Case study:* The TX Power reference is still fixed in the previous experiment, we now conduct a case study to figure out how it affects the localization result. By using different TX power references of the three fake iBeacons, we record all possible estimated locations under attack, as shown in Fig. 23. Since the estimation algorithm does not provide an extremely close distance, we see very few attacked positions nearing iBeacons. For other places, fake positions nearly cover the whole area. This indicates a high ability to mislead the user.

### E. Fingerprint Localization

In this section, we conduct attacks on fingerprint-based localization. We place iBeacons and fake iBeacons in an office building shown in Fig. 16. 7 iBeacons are used to build the

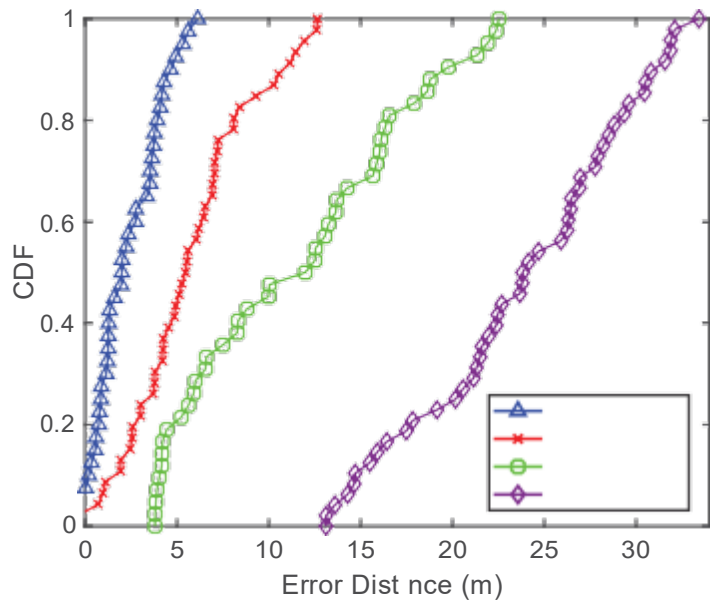

Fig. 24: Distance error of AP number with fingerprint attack

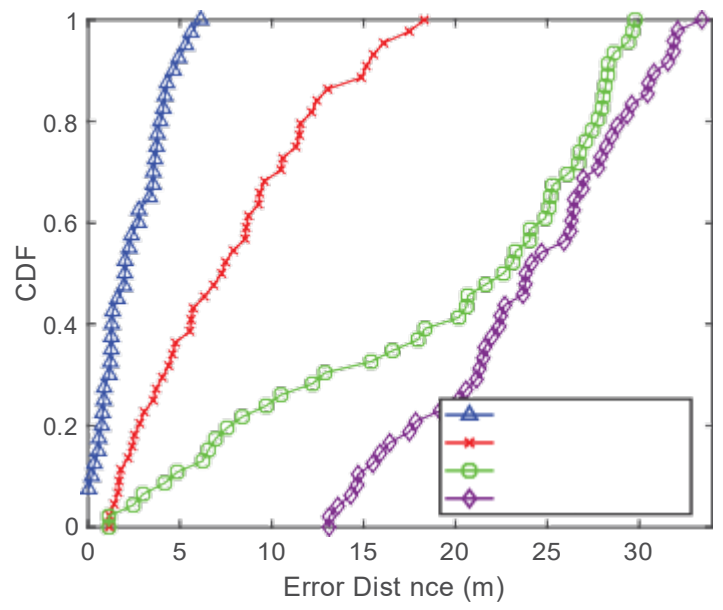

Fig. 25: Distance error of using one AP to impersonate multiple iBeacons

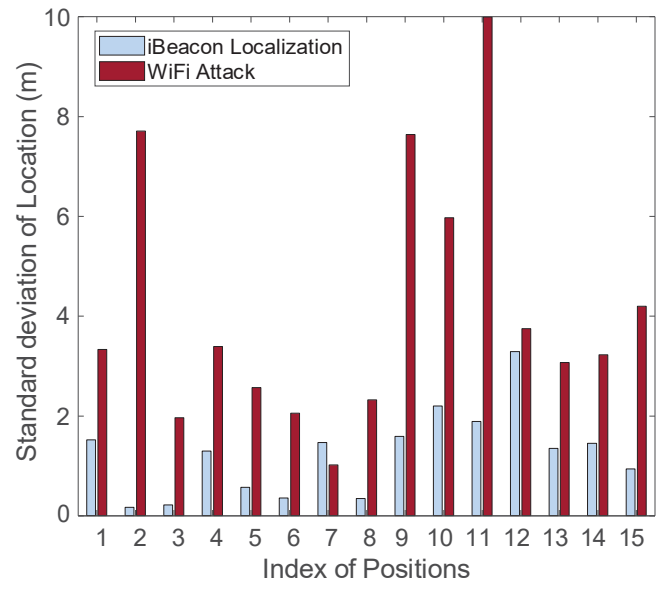

Fig. 26: Impact of multipath propagation on localization result

RSSI database and 6 WiFi APs are used to conduct attacks. We collect the RSSI fingerprint and locations of 120 spots.

*1) Attack performance:* Firstly, We show the performance of using different numbers of APs to conduct the attack. The distance error result is shown in Fig. 24. If only one or two WiFi APs are used, the average distance error will only raise from 2.7 meters to 5.2 meters. This is because the system is deployed in a large area and the effective range is small with few fake iBeacons. By adding more APs to attack, the effective range gradually expands to the whole area and the error increases significantly.

*2) Multiple impersonation performance:* In this experiment, we evaluate how multiple impersonation benefits the attack. We let three fake iBeacons broadcast two iBeacon packets each. In this way, we obtain 6 virtual fake iBeacons. We compare the distance error by using 6 true APs. The result is shown in Fig. 25. With the ability to impersonate more than one iBeacon, the estimation error greatly increases at the places where the attacker is able to affect. But for the spots out of the effective range of the 3 fake iBeacons, the error is still not very high. However, the error we provide is already enough to invalidate the localization system.

*3) Impact of multipath propagation:* Compared with iBeacon, WiFi AP provides a much more unstable RSSI measurement in the same environment. In this experiment, we verify the effect of more multipath propagation in WiFi. We select 15 spots in the area and record their RSSI for 20 seconds. The standard variation of the estimated location is shown in Fig. 26. It is clear that more multipath propagation of WiFi APs brings more uncertainty during localization, and therefore benefits the attack.

## VI. Related Work

### *A. Attacks and Countermeasures on RSS-IPS*

RSS-based indoor positioning system has been studied extensively during the last two decades. Most of the work focus on how to match the measured RSSI fingerprint in the database to acquire higher localization accuracy [20], [21]. Many works have also been proposed to attack these indoor positioning systems (IPS). The authors in [22] formulate all-around signal strength attack where similar attacks are launched at all transmitters. In [19], false data injection attack in the crowdsourced IPS system is presented. More recently, the authors in [9] study impersonation attacks on WiFi-based IPS systems. They present algorithms to achieve specific attack aims and provide countermeasures for their attacks. These algorithms can all be applied using Wi-attack.

To defend from attacks, the authors in [23] use K-means clustering to distinguish "good" APs from the attacked ones based on geometric relationships between estimated results. Similarly, the authors in [24] propose to localize using selected reliable APs based on their scores. In [25], the authors propose a voting method to improve robustness where all WiFi APs give a vote on the reference location after receiving the user's RSS measurement. However, existing countermeasures encounter difficulties when one attacker AP is able to impersonate multiple transmitters.

### *B. Cross Technology Communication and attacks*

Cross-technology communication enables direct communication for heterogeneous wireless protocols and has been applying to hybrid networking [26], [27]. Early works utilize packet-level information to communicate, such as broadcasting intervals [28], signal strength [11], or channel state information [29], [30]. These methods often provide low throughput. To improve it, WEBee [12] proposes direct analog signal emulation from WiFi to ZigBee, which is then enhanced by WIDE [13] using digital emulation. However, none of these works can be used on WiFi to BLE communication. In this work, we propose to split the phase ladder in WIDE into fine-grained ones and transmit with an enhanced method. The result in Section V-B shows that Wi-attack increases the PRR significantly. We notice that [31] also enables WiFi-BLE CTC, but it is implemented based on 802.11b. It can not be extended to up-to-date WiFi standards (e.g., 802.11g/n/ac).

There are also works that conduct attacks between different technologies or provides countermeasures. The authors in [16] show that it is easy to conduct jamming attacks from WiFi to ZigBee by simply adjusting transmission parameters. They propose the method to decode ZigBee packets while being jammed. In [32], the authors attack ZigBee nodes by transmitting control packets from WiFi. To defend it, an identifying method is proposed which distinguishes emulated packets from normal packets by constellation diagram analysis.

## VII. Discussion

In this section, we discuss the possible ways to detect and defend against Wi-attack. There are two possible ways of detecting the existence of Wi-attack. First, if the attacker raises the transmission power as suggested in point deployment to an abnormal level, emulated packets can be filtered easily based on the RSSI value. Therefore, in order to hide from user detection, the attacker should carefully select the transmission power and put the RSSI within a normal range. Second, considering that Wi-attack provides only 70% PRR in total, it is possible to detect it by monitoring the received packet number over time. However, we argue that this is difficult in practice. On one hand, since the user may be moving, localization should be done instantly after receiving the packet. It is demanding and wasting to require a long monitoring phase before each positioning. On the other hand, even if a lower PRR is detected, the application cannot judge if any of the received packets is from an attacker's AP, since the AP is transmitting the same iBeacon packet. The same situation could occur if the user is far away from the Beacon or the channel condition is bad. Therefore, detecting and defending Wi-attack remains to be difficult in a real-world situation.

## VIII. Conclusion

This work reveals the feasibility of conducting impersonation attacks on iBeacon services using commercial off-the-shelf WiFi devices. By emulating redundant iBeacon advertisements, Wibeacon provides a high packet reception rate (up to 66.2%) and stability in order to launch effective attacks. We consider three typical scenarios and propose corresponding attack methods. To evaluate Wi-attack, we implement point deployment, multilateral localization, and fingerprint-based localization systems on the real-world deployment. The result of our attack method shows that Wi-attack conducts a highly effective attack on all three deployments. Especially, Wi-attack can bring an average distance error of more than 20 meters on fingerprint-based localization, which can invalidate such systems with extremely low overhead.

## Acknowledgment

This work is supported in part by National Key R&D Program of China No. 2017YFB1003000, National Science Fund of China under grant No. 61772306, and the R&D Project of Key Core Technology and Generic Technology in Shanxi Province (2020XXX007).

## References

[1] Y. Michalevsky, S. Nath, and J. Liu, "Mashable: Mobile applications of secret handshakes over bluetooth le," in *Proceedings of ACM Mobicom*, 2016.

[2] P. Kindt, D. Yunge, M. Gopp, and S. Chakraborty, "Adaptive online power-management for bluetooth low energy," in *Proceedings of IEEE INFOCOM*, 2015.

[3] P. Kriz, F. Maly, and T. Kozel, "Improving indoor localization using bluetooth low energy beacons," *Mobile Information Systems*, 2016.

[4] G. Solmaz, J. Fürst, S. Aytaç, and F.-J. Wu, "Group-in: Group inference from wireless traces of mobile devices," in *Proceedings of ACM/IEEE IPSN*, 2020.

[5] "Apple ibeacon," 2013. [Online]. Available: https://developer.apple.com/ibeacon/

[6] "Smartphones deals turn store shoppers into buyers," 2015. [Online]. Available: https://www.timesunion.com/tuplus-local/article/Smartphones-deals-turn-store-shoppers-into-buyers-6693260.php

[7] C. Jiang, Y. He, S. Yang, J. Guo, and Y. Liu, "3d-omnitrack: 3d tracking with cots rfid systems," in *Proceedings of ACM/IEEE IPSN*, 2019.

[8] W. Wang, J. Li, Y. He, and Y. Liu, "Symphony: Localizing multiple acoustic sources with a single microphone array," in *Proceedings of ACM Sensys*, 2020.

[9] L. Yuan, Y. Hu, Y. Li, R. Zhang, Y. Zhang, and T. Hedgpeth, "Secure rss-fingerprint-based indoor positioning: Attacks and countermeasures," in *Proceedings of IEEE CNS*, 2018.

[10] "Beecon," 2016. [Online]. Available: http://www.beaconsandwich.com/

[11] X. Guo, X. Zheng, and Y. He, "Wizig: Cross-technology energy communication over a noisy channel," in *Proceedings of IEEE INFOCOM*, 2017.

[12] Z. Li and T. He, "Webee: Physical-layer cross-technology communication via emulation," in *Proceedings of ACM MobiCom*, 2017.

[13] X. Guo, Y. He, J. Zhang, and H. Jiang, "Wide: Physical-level ctc via digital emulation," in *Proceedings of ACM/IEEE IPSN*, 2019.

[14] M. K. Hanawal, D. N. Nguyen, and M. Krunz, "Jamming attack on in-band full-duplex communications: Detection and countermeasures," in *Proceedings of IEEE INFOCOM*, 2016.

[15] L. Tang, Y. Sun, O. Gurewitz, and D. B. Johnson, "Em-mac: A dynamic multichannel energy-efficient mac protocol for wireless sensor networks," in *Proceedings of ACM MobiHoc*, 2011.

[16] Z. Chi, Y. Li, X. Liu, W. Wang, Y. Yao, T. Zhu, and Y. Zhang, "Countering cross-technology jamming attack," in *Proceedings of ACM WiSec*, 2020.

[17] F. Xu, Z. Qin, C. C. Tan, B. Wang, and Q. Li, "Imdguard: Securing implantable medical devices with the external wearable guardian," in *Proceedings of IEEE INFOCOM*, 2011.

[18] M. Jin, Y. He, X. Meng, D. Fang, and X. Chen, "Parallel backscatter in the wild: When burstiness and randomness play with you," in *Proceedings of ACM Mobicom*, 2018.

[19] T. Li, Y. Chen, R. Zhang, Y. Zhang, and T. Hedgpeth, "Secure crowdsourced indoor positioning systems," in *Proceedings of IEEE INFOCOM*, 2018.

[20] P. Bahl and V. N. Padmanabhan, "Radar: An in-building rf-based user location and tracking system," in *Proceedings of IEEE INFOCOM*, 2000.

[21] M. Youssef and A. Agrawala, "The horus wlan location determination system," in *Proceedings of ACM Mobisys*, 2005.

[22] X. Li, Y. Chen, J. Yang, and X. Zheng, "Designing localization algorithms robust to signal strength attacks," in *Proceedings of IEEE INFOCOM*, 2011.

[23] J. Yang, Y. Chen, V. B. Lawrence, and V. Swaminathan, "Robust wireless localization to attacks on access points," in *IEEE Sarnoff Symposium*, 2009.

[24] A. Kushki, K. N. Plataniotis, and A. N. Venetsanopoulos, "Sensor selection for mitigation of rss-based attacks in wireless local area network positioning," in *Proceedings of IEEE ICASSP*, 2008.

[25] Y. Li, Y. Hu, R. Zhang, Y. Zhang, and T. Hedgpeth, "Secure indoor positioning against signal strength attacks via optimized multi-voting," in *Proceedings of ACM IWQoS*, 2019.

[26] J. Zhang, X. Guo, H. Jiang, X. Zheng, and Y. He, "Link quality estimation of cross-technology communication," in *Proceedings of IEEE INFOCOM*, 2020.

[27] Z. Yu, P. Li, C. A. Boano, Y. He, M. Jin, X. Guo, and X. Zheng, "Bicord: Bidirectional coordination among coexisting wireless devices," in *Proceedings of IEEE ICDCS*, 2021.

[28] S. M. Kim and T. He, "Freebee: Cross-technology communication via free side-channel," in *Proceedings of ACM MobiCom*, 2015.

[29] X. Guo, Y. He, X. Zheng, L. Yu, and O. Gnawali, "Zigfi: Harnessing channel state information for cross-technology communication," in *Proceedings of IEEE INFOCOM*, 2018.

[30] W. Wang, X. Zheng, Y. He, and X. Guo, "Adacomm: Tracing channel dynamics for reliable cross-technology communication," in *Proceedings of IEEE SECON*, 2019.

[31] L. Li, Y. Chen, and Z. Li, "Poster abstract: Physical-layer cross-technology communication with narrow-band decoding," in *Proceedings of IEEE ICNP*, 2019.

[32] X. Zhang, P. Huang, L. Guo, and Y. Fang, "Hide and seek: Waveform emulation attack and defense in cross-technology communication," in *Proceedings of IEEE ICDCS*, 2019.